\documentclass[universe,communication,accept,moreauthors,pdftex]{Definitions/mdpi} 
\makeatletter
\let\c@lofdepth\relax
\let\c@lotdepth\relax
\makeatother

\usepackage{subfigure}
\makeatletter
\renewcommand{\@thesubfigure}{\small(\textbf{\alph{subfigure}})}
\makeatother

\firstpage{1} 
\pubvolume{1}
\issuenum{1}
\articlenumber{0}
\pubyear{2023}
\copyrightyear{2023}
\externaleditor{{Academic Editor: Lorenzo Iorio} 

} 
\datereceived{20 March 2023} 
\daterevised{27 June 2023 } 
\dateaccepted{14 July 2023} 
\datepublished{ } 
\hreflink{https://doi.org/} 

\Title{A Generalized Double Chaplygin Model for Anisotropic Matter: The Newtonian Case}
\TitleCitation{A Generalized Double Chaplygin Model for Anisotropic Matter: The Newtonian Case}

\Author{
Gabriel Abell\'an 
 $^{1,\dagger}$\orcidA{}, 
\'Angel Rinc\'on $^{2,3}$*$^{,\dagger}$\orcidB{}
and
Eduard Sanchez $^{1,\dagger}$,
}

\AuthorNames{Firstname Lastname, Firstname Lastname and Firstname Lastname}

\AuthorCitation{Abell\'an, G.; Rinc\'on, \'A.; Sanchez, E.}

\address{%
$^{1}$ \quad Escuela de F\'isica, Facultad de Ciencias, Universidad Central de Venezuela, Caracas 1050, Venezuela; 
gabriel.abellan@ciens.ucv.ve (G.A.); eduard.sanchez@ucv.ve (E.S.) 
\\
$^{2}$ \quad Departamento de F{\'i}sica Aplicada, Universidad de Alicante, Campus de San Vicente del Raspeig, \mbox{E-03690 Alicante, Spain}
\\
$^{3}$ \quad Sede Esmeralda, Universidad de Tarapac{\'a}, Avda. Luis Emilio Recabarren 2477, Iquique, Chile.
}

\corres{Correspondence:  angel.rincon@ua.es}

\firstnote{These authors contributed equally to this work.}

\abstract{In this work, we investigate astrophysical systems in a Newtonian regime using anisotropic matter. For this purpose, we considered that both radial and tangential pressures satisfy a generalized Chaplygin-type equation of state. Using this model, we found the Lane--Emden equation for this system and solved it numerically for several sets of parameters. Finally, we explored the mass supported by this physical system and compared it with the Chandrasekhar mass.
}


\begin{document}



\section{Introduction}\label{Intro}
The Newtonian theory of the stellar structure is typically used to describe stars with low densities and weak gravitational fields \cite{Chandrasekhar1939,Horedt2004,Kippenhahn2012}. Indeed, the application of Newtonian gravity, even for compact, high-density objects such as white dwarfs and neutron stars, can sometimes provide acceptable results that are comparable to those of relativistic models~\cite{Rodrigues_2013}.
It is essential to point out that relativistic solutions (even including pressure) can be mimicked by using neo-Newtonian hydrodynamics \cite{Fabris:2014ena}. The latter theory was conceived to extend the traditional Newtonian theory to include the usual relativistic effects while keeping the simplicity of the well-known Newtonian framework. Thus, we have different levels to face the problem of stellar distributions:
(i) standard Newtonian theory, 
(ii) neo-Newtonian theory 
and 
(iii) general relativity. 

In all the above-mentioned cases, equations of state are useful tools for making progress, allowing us to gain insights about certain properties of interior solutions \cite{Oertel:2016bki}. In particular, the selection of a concrete equation of state of stellar interiors has a profound impact on the inner properties of the star. The reason for this is that an equation of state encodes the microscopic properties of stellar matter (for a given density $\rho$, temperature $T$ and composition $X_i$). %
Also, by taking advantage of the laws of thermodynamics and a similar equation for the internal energy $U(\rho, T, X_i)$, it is possible to derive from the equation of state the thermodynamic properties required to describe the structure of a star (such as the specific heats $c_V$ and $c_P$ \cite{Nilsson:2000zg}).

Special attention should be dedicated to a particular kind of equation: the polytropic equation of state. The latter has played a crucial role  (see \cite{Abramowicz1983,Chandrasekhar1939,Horedt2004,Kippenhahn2012,Singh:2020cnu} and the references therein). A polytropic equation of state is simple, and it allows closing of the system to be solved 
\cite{Mardan:2020noh,Singh:2020cnu,Kumar:2022kho,Azam:2016eek,Azam:2016qua,Ngubelanga:2015ith,Leon:2023yvz,Ovalle:2022yjl,Ramos:2021drk,2000IJMPD...9...35J}. (Some generalizations were investigated in \cite{Azam:2016qua,Azam:2016eek}.) 
Furthermore, the Lane--Emden equation that describes the system takes advantage of a polytropic equation of state to provide the characteristics of the density profile \cite{Mach:2012dj,Suleiman:2022egw,Santana:2022vmw,LasHeras:2022pyj}.

Thus, in order to gain insights into the underlying physics behind compact stars, it is always interesting to investigate how some non-trivial equations of state can alter the dynamics of the physical system \cite{Khunt:2021dpi}. 
While we restrict ourselves to the Newtonian case in this work, it should be stated that for extremely compact objects, a relativistic theory of gravitation (e.g., general relativity) must be applied, which demands its own adapted approach. What is more, if a detailed relativist treatment is not included, then neo-Newtonian theory could serve to make progress.

In most applications, it is quite common to assume that the fluid distribution satisfies Pascal's principle (equal principal stresses), meaning that the pressure is assumed to be isotropic. Nowadays, however, we know that the isotropic pressure condition may be too severe, and furthermore, the presence of local anisotropy may be caused by a wide variety of physical phenomena that are expected to be found in compact objects \cite{Herrera:1997plx}, such as (1) some velocity distribution in collisionless gas \cite{Binney:2008a}, (2) rotational effects \cite{Herrera:1997plx,Horedt2004}, (3) mixtures of two or more perfect fluids \cite{Alencar:1986a} and (4) the repulsive electrostatic force in charged systems \cite{Liu:2014jna}. 
Hence, as previously mentioned, there exist various sources of anisotropy, such as those arising from anisotropic velocity distributions \cite{1922MNRAS..82..122J}. However, it should be noted that the effect of anisotropy may diminish in the non-relativistic limit, as is the case when the pressure anisotropy is attributed to magnetic fields present at the cores of compact stars~\cite{Yagi:2015hda}.

Recently, some models of anisotropy have been proposed under the assumption that the tangential pressure also obeys a polytropic-like equation of state \cite{Abellan:2020ceu,Leon:2021ncw,Azam:2022qqj}. These models enable different cases to be particularized by adjusting the parameters. In this work, we are interested in extending these polytropic models by considering a generalized Chaplygin equation of state \cite{Bertolami:2005pz,Tello-Ortiz:2020svg,Gorini:2008zj}. 
In this case, the radial and tangential pressures adopt the following form:
\begin{eqnarray}\label{genChap}
        p = K\rho^\gamma - \frac{M}{\rho^N}\;\;, \hspace{1cm}
        \gamma = 1 + \frac{1}{n}\;.
\end{eqnarray}
In 
this equation, $p$ and $\rho$ denote the pressure and density, respectively. On the other hand, $K$ is the polytropic constant, $M$ is the Chaplygin constant, $\gamma$ is the polytropic exponent, $n$ is the polytropic index, and $N$ is the Chaplygin index. 
Please notice that a generalized Chaplygin equation of state could generate a Chaplygin dark star (in the scenario of unification of dark energy and dark matter), and such a type of hypothetical object still needs more study. Thus, as this kind of astrophysical phenomenon needs this particular type of equation of state, we are forced to consider equations of state like Equation \eqref{genChap} to make progress. 
In addition, the stability of compact objects relies on preventing the gravitational collapse of their mass, which can be achieved through several conditions such as hydrostatic force and Coulomb's force from the electric charge. In this respect, the generalized Chaplygin equation of state (EOS) can be a useful tool for incorporating new theories, such as the dark fluid model and dark energy stars (as we previously said). These theories introduce an additional repulsive force to the compact star model, which can enhance its stability \cite{Prasad:2021hrx}. 
It is also crucial to note that there exist various constraints on the parameter space of the equation of state (EoS), including those derived from a cosmological perspective using astronomical data~\cite{2012EPJC...72.1931X}. These constraints provide valuable insights and contribute to our understanding of the EoS and its implications for cosmology.

A theory of Newtonian polytropes for anisotropic matter was fully developed in \cite{Herrera:2013dfa}. (For the relativistic version, see \cite{Herrera:2013fja}.) Here, we will closely follow the approach outlined in \cite{Shojai:2015hsa}, assuming this time that both pressures (radial and tangential) satisfy an extended Chaplygin equation of state.
Furthermore, it should be noted that our work is distinct from previously published research on the generalized polytropic equation of state. To date, no study has been conducted on the double anisotropic case, making our findings particularly novel. 
By doing so, the resulting Lane--Emden equation can be integrated, and the models would depend on the specific parameters involved. It is essential to point out that such equations of state have been significantly investigated, on the one hand, in the cosmological context \cite{Pourhassan:2013sw,Panotopoulos:2020qbx} and, on the other hand, in the physics of relativistic compact stars \cite{Panotopoulos:2020kgl,Panotopoulos:2021dtu}.

This work is organized as follows. In Section~\ref{sec:2}, we outline the general method for treating Newtonian generalized Chaplygin models for anisotropic matter. In order to integrate the resulting Lane--Emden equation, we need an additional ansatz. We assume that both the radial and tangential pressures obey a generalized Chaplygin equation of state like Equation \eqref{genChap}. This is carried out in Section \ref{sec:3}. Next, in Section \ref{sec:4}, the Lane--Emden equation is integrated numerically for specific values of the parameters. Finally, a discussion of the results is presented in the last section.

\section{Anisotropic Generalized Chaplygin}\label{sec:2}
In this section, we will briefly introduce the main ingredients needed to deal with anisotropic matter, taking advantage of a concrete type of equation of state. This idea is inspired by a previous work \cite{Herrera:2013dfa} and extended for a generalized Chaplygin equation of state. 
First, it should be mentioned that polytropes (self-gravitating gaseous spheres that are used as a first approximation to face more realistic stellar models) help us obtain insights into the physics of a star, although in most cases, it is possible via numerical integration. Polytropes offer a more tractable way to estimate various internal quantities of the star. A quite natural extension of the polytropic equation of state is the well-known Chaplygin equation. The latter can be understood as a polytropic equation of state plus an additional term inversely proportional to density. This expression has two free parameters, which make such an equation versatile. In particular, an extended Chaplygin fluid offers a way to apply this type of model to a broader class of physical systems.
In the presence of anisotropic matter, and when its contribution is only diagonal, we can write the hydrostatic equilibrium equation in terms of spherical coordinates as follows:
\begin{equation}\label{sec2_01}
    \frac{dP_r}{dr} \;=\; -\rho \frac{d\phi}{dr} + \frac{2}{r}\Delta\; ,
\end{equation}
where $\phi$ is the gravitational potential and $\Delta \equiv P_\perp - P_r$ is the anisotropy factor. The last equation is the Newtonian version of the well-known Tolman--Opphenheimer--Volkoff equation in the presence of anisotropic matter. Notice that the reason for only two different principal stresses (i.e.,  $P_\theta = P_\varphi = P_\perp$ and $P_r \neq P_\perp$) is a consequence of spherical symmetry.
Allow us to provide further elaboration on the anisotropic hydrostatic equilibrium equation. Generally, there are several approaches to deducing Equation \eqref{sec2_01}, including
  (1)  a geometrical procedure,
(2)  an analytical procedure and 
(3) a limit of general relativity.
The geometric derivation assumes a thin mass element in a star with a thickness $dr$ and surface $dA$ at a radius $r$ from the center and applies the principle of hydrostatic equilibrium while considering three contributions: 
(1) $dF_g$, which represents the force due to gravity,
(2) $dF_{P_r}$, denoting the force due to radial pressure, and 
(3) $dF_{P_{\perp}}$, representing the force due to tangential pressure.
Moreover, the tangential contribution can be further divided into orthogonal components, with the radial component being the pertinent one. To sum up these three contributions, we arrive at the following expression:
\begin{align}
4 \pi r^2 dr (d\theta)^2 P_r + 2 \pi r^2 dr (d\theta)^2 \frac{dP_r}{dr} - 4 \pi r^2 dr (d\theta)^2 P_{\perp} &= -\rho \frac{G m(r)}{r^2} 2 \pi r^2 dr (d\theta)^2.
\end{align}
By simplifying and rewriting the last equation, we obtain Equation \eqref{sec2_01}.
The analytical derivation simply involves the direct application of the fundamental equation of hydrostatics. Similar to what occurs in elastostatics for solids, we can utilize the following equation in hydrostatics:
\begin{align} \label{asdf}
\vec{\nabla} \cdot \overleftrightarrow{T} &= -\vec{f},
\end{align}
where $\overleftrightarrow{T}$ is the fluid's stress tensor and $\vec{f}$ is the external force density. While assuming a diagonal anisotropic stress tensor and radial dependence only, we have 
\begin{align}
\overleftrightarrow{T} &= \text{diag}\Bigl( P_r, P_{\perp}, P_{\perp} \Bigl). 
\end{align}
With the latter in mind, we can evaluate Equation \eqref{asdf} to obtain
\begin{align}
\frac{dT_{rr}}{dr} + \frac{1}{r} \Bigl( 2T_{rr} - T_{\theta \theta} - T_{\phi \phi}  \Bigl) &= -f.
\end{align}
To complete this second (alternative) derivation, we need to remember that $T_{rr} \equiv P_r$, $T_{\theta \theta} \equiv P_{\perp} = T_{\phi \phi}$ and $f$ (the radial part) is given by
\begin{align}
f &= -\rho \frac{G m(r)}{r^2} ,
\end{align}
to finally recover Equation \eqref{sec2_01}. The last possibility takes advantage of Einstein's field equations. After calculating the corresponding Tolman--Oppenheimer--Volkoff (TOV) equation, we arrive at the simplified equation (with $G=1$):
\begin{align}
\frac{dP_r}{dr} &= -(\rho + P_r)\Bigg(\frac{m(r)+4 \pi r^3 P_r}{r(r-2m)}\Bigg) + \frac{2}{r}\Bigl( P_{\perp} - P_r \Bigl),
\end{align}
where it is necessary to assume the following:
\begin{align}
\frac{m(r)}{r} &<< 1 , 
\\
4 \pi r^3 P_r &<< m(r),
\\
P_r &<< \rho.
\end{align}
Finally, including the last assumptions, we obtain the standard equation:
\begin{align}
\frac{dP_r}{dr} &= -\rho \frac{m(r)}{r^2} + \frac{2}{r}\Bigl( P_{\perp} - P_r \Bigl).
\end{align}
All these alternative approaches can be consulted in \cite{1974ApJ...188..657B,Dev:2003qd,1989fsa..book.....C,Herrera:1997plx}. 
The behavior of the gravitational potential $\phi$ is determined via the Poisson equation
\begin{equation}\label{sec2_02}
    \frac{1}{r^2} \frac{d}{dr}\left( r^2 \frac{d\phi}{dr} \right)
    \; =\; 4\pi G\rho\;,
\end{equation}
with $G$ being Newton's constant. For the radial pressure, we will assume a generalized version of the Chaplygin equation of state (Equation \eqref{genChap}). In terms of the standard parametrization $\rho = \rho_c \omega^{n_r}$, we could write without loss of generality 
\begin{align} \label{pradial}
        P_{r} = P_{r0}\left[\omega^{1+n_{r}} - \frac{\mathcal{M}_{r}}{\omega^{N_{r}}}\right] .
\end{align}
In this case, $P_{r0}$ is a constant related to the pressure at the center of matter distribution, $n_r$ is the index of the radial polytrope, $N_r$ is the Chaplygin index, and $\mathcal{M}_r$ is the radial Chaplygin constant. An essential consequence of Equation \eqref{pradial} is that, in general, $\omega \neq 0$ at the surface $\Sigma$ of the spherical configuration. In particular, we find that
\begin{equation} \label{p-cond}
    [P_r]_\Sigma = 0 \hspace{.4cm} \longrightarrow \hspace{.4cm}
    \omega_\Sigma = (\mathcal{M}_r)^{1/1+n_r+N_r}\,.
\end{equation}
This condition reduces to the well-known polytropic condition $\omega_\Sigma = 0$ when $\mathcal{M}_r = 0$.

By using the derivative of Equation \eqref{pradial} in Equation \eqref{sec2_01} and substituting in the Poisson equation (Equation \eqref{sec2_02}), we find the Lane--Emden equation for the generalized Chaplygin system:

\begin{align} \label{lanenew}
    \alpha \omega'' +
    \beta  \omega' + 
    \gamma = -\omega^{n_r}
\end{align}
where the primes denote differentiation with respect to the coordinate $z$ and the coefficients $\alpha \equiv \alpha(\omega, z)$, $\beta \equiv \beta(\omega, z)$ and $\gamma \equiv \gamma(\omega, z)$ are given by
\begin{align}
    \alpha &= 1+\frac{N_{r}\mathcal{M}_{r}}{(1+n_{r})\omega^{1 + n_{r}+N_{r}}} , \label{alpha}
    \\
    \beta &=  \frac{2}{z}+\frac{N_{r}\mathcal{M}_{r}}{(1+n_{r})\omega^{1+n_{r}+N_{r}}}  \left[ \frac{2}{z}-\frac{1+n_{r}+N_{r}}{\omega} \omega' \right] ,
    \\
    \gamma &=    -\frac{2}{P_{r0}(1+n_{r})z\omega^{n_{r}}} \left( \Delta' + \frac{\Delta}{z} - n_{r} \frac{\omega'}{\omega}  \Delta   \right) .
\end{align}
Here, we have used the following redefinition:
\begin{align}
    z &= A r\;, \label{zzz}
    \\
    A^{2} &= \frac{P_{r0}(1+n_{r})}{4 \pi G \rho^{2}_c} \;.
    \label{AAA}
\end{align}
Note that when $\mathcal{M}_r \rightarrow 0$, we recover the simplest case previously obtained in \cite{Shojai:2015hsa}. 
Also note that at the center of the system, we have $\omega(0) = 1$. The isotropic limit $\Delta\to 0$ produces the well-known Lane--Emden equation \cite{Kippenhahn2012,Herrera:2013dfa}. 
Equation \eqref{lanenew} is a second-order differential equation, and therefore we must give two conditions in order to determine the complete~solution.

In order to make progress on the condition on the first derivative of $\omega$, we will take advantage of the hydrostatic equilibrium and the Poisson equation. 
The equation of hydrostatic equilibrium was previously introduced in Equation \eqref{sec2_01}. What is more, if we replace the definitions in that equation, we have
\begin{align} \label{proeq}
     P_{r0}\left[(1+n_{r})\omega^{n_{r}}+\frac{N_{r}\mathcal{M}_{r}}{\omega^{1+N_{r}}}  \right] &= -\rho_{c} \, \omega^{n_{r}} \frac{d \phi}{d r} + \frac{2}{r} \Delta\;.
\end{align}
After that, we utilize the first integral of the Poisson equation, namely
\begin{align} \label{partialphi}
    \frac{d \phi}{d r} &= \frac{4 \pi G \rho_{c}}{r^{2}} \int^{r}_{0} \Tilde{r}^{2} \omega^{n_{r}} d\Tilde{r} \;.
\end{align}
By placing Equation \eqref{partialphi} into Equation \eqref{proeq} and taking into account the rescaling given by Equations \eqref{zzz} and \eqref{AAA}, we obtain
\begin{align}\label{omegaprime}
\begin{split}
       \omega'= \frac{
       \left[  - (1+n_{r})\frac{\omega^{n_{r}}}{z^{2}} \int^{z}_{0} x^{2} \omega^{n_{r}} \, dx  +  \frac{2}{P_{r0} \, z} \Delta   \right]
       }
       {
       \left[ (1+n_{r})\omega^{n_{r}}+\frac{N_{r}\mathcal{M}_{r}}{\omega^{1+N_{r}}}   \right] 
       }.
\end{split}
\end{align}
The above expression is reduced to those obtained in \cite{Abellan:2020ceu} under the limit $N_r \rightarrow 0$. It should be noticed that when $z=0$, we obtain 
 \begin{align}
     \frac{1}{(1+n_{r})\omega^{n_{r}}+\frac{N_{r}\mathcal{M}_{r}}{\omega^{1+N_{r}}}}     
     \hspace{.3cm} \rightarrow \hspace{.3cm} 
     \frac{1}{(1+n_r) + N_r \mathcal{M}_r}\;,
 \end{align}
where we have used the condition $\omega(0)=1$. Note that in the numerator of Equation \eqref{omegaprime}, we have an indetermination when $z\to 0$. By using the L'Hopital rule on the first term, we obtain 
\begin{align}
    \frac{\omega^{n_{r}}}{z^{2}} \int_{0}^{z} z^{2} \omega^{n_{r}} \, dz \hspace{.4cm} \xrightarrow[]{x\to 0} \hspace{.4cm}
    \frac{1}{2} z \, \omega^{2n_{r}} = 0\;.
\end{align}
Finally, we have
 \begin{equation}
     \omega'(0)= \left( \frac{1}{(1+n_{r})+N_{r}\mathcal{M}_{r}}
     \right)
     \frac{2}{P_{r0}} \lim_{z \to 0} \frac{\Delta}{z}.
 \end{equation}
In addition, supplementary information is required to integrate the Lane--Emden equation (Equation~\eqref{lanenew}) and obtain specific solutions. Thus, to close the system, we impose a condition on the tangential anisotropy. In the next sections, we will introduce the general form of the anisotropy factor $\Delta$ for the double Chaplygin generalized model. At the same time, we can complete the analysis of the appropriate condition for $\omega'$ at the origin.

\section{The Generalized Double Chaplygin Model}\label{sec:3}

From now on, we will consider that the two existent pressures (radial and tangential) are parameterized by a generalized Chaplygin equation of state:
\begin{align}
    P_{r} &= P_{r0}\left[\omega^{1+n_{r}} - \frac{\mathcal{M}_{r}}{\omega^{N_{r}}}\right], 
    \\
    P_{\perp} &= P_{\perp 0}\left[ \omega^{n_{r}} \omega^{\theta}-\frac{\mathcal{M}_{\perp}}{\omega^{N_{\perp}}}  \right].
\end{align} 
At this point, it is important to emphasize that, as is often the case in related scenarios involving anisotropic stars, we require auxiliary information to obtain concrete models. In this instance, an auxiliary assumption can be made for the transverse pressure similar to that in the radial case. This approach is particularly applicable to weakly anisotropic distributions (i.e., small $\Delta$). Under this assumption, we can connect our solution smoothly by varying the parameters (including the isotropic pressure case) in a natural way. While a generalized Chaplygin equation of state has been used previously in the context of relativistic stars \cite{Bhar:2016nlq,Estevez-Delgado:2021phh,Bhar:2016lfu,Estevez-Delgado:2021cgb,Rincon:2023ens}, the generalized {\it{double}} Chaplygin equation of state is a novel and unexplored approach that deserves considerable attention.
Using these expressions, we can write the anisotropy factor ($ \Delta = P_{\perp}-P_{r}$) as follows:
\begin{align}
    \Delta &= P_{\perp 0}\left[ \omega^{n_{r}} \omega^{\theta}-\frac{\mathcal{M}_{\perp}}{\omega^{N_{\perp}}}  \right]-P_{r0}    
    \left[\omega^{1+n_{r}}-\frac{\mathcal{M}_{r}}{\omega^{N_{r}}}\right].
\end{align}
Knowing that at the center of the matter distribution, we have $\Delta(0) = 0$, we find the following constraint between parameters:
\begin{equation}\label{constraint}
    \frac{P_{\perp 0}}{P_{r0}} = 
    \frac{1 - \mathcal{M}_r}{1 - \mathcal{M}_\perp}\,.
\end{equation}
With this condition, the anisotropy factor is expressed as follows:
\begin{align}
\begin{split} \label{deltafactor}
    \Delta = P_{r 0}\Bigg[
    &\left(\frac{1-\mathcal{M}_{r}}{1-\mathcal{M}_{\perp}}\right) \left(\omega^{n_{r}} \omega^{\theta}-\frac{\mathcal{M}_{\perp}}{\omega^{N_{\perp}}}\right)  -
    \left(\omega^{1+n_{r}}-\frac{\mathcal{M}_{r}}{\omega^{N_{r}}}\right)
    \Bigg].
    \end{split}
\end{align}
Note that in order to obtain the isotropic case, we must consider the following three limits:
\begin{eqnarray}
    \theta &\to& 1 \,, \\
    \mathcal{M}_\perp &\to& \mathcal{M}_r \,,\\
    N_\perp &\to& N_r \,.
\end{eqnarray}

In the next section, we will introduce the corresponding Lane--Emden equation in terms of the generalized Chaplygin equation of state.

\section{Solving the Anisotropic Lane--Emden Equation}\label{sec:4}
As was pointed out before, we are investigating the Lane--Emden equation for particular cases of the anisotropy factor $\Delta$. Thus, we consider three well-defined cases which are limit cases of the general expression in Equation \eqref{deltafactor}.
%
\subsection{Case 1}
%
First, we study the case where $\mathcal{M}_{\perp} = \mathcal{M}_r$ and $N_{\perp} = N_r$ to obtain the corresponding anisotropic factor as follows:
\begin{align}
    \Delta = P_{r0} \, \omega^{n_{r}}(\omega^{\theta} - \omega).
\end{align}

Although a complete analytical solution is not possible to achieve, we can accomplish this numerically. However, we still can verify the behavior of $\omega'(0)$. Using Equation \eqref{omegaprime}, we obtain
\begin{align}
    \omega'(0)=\frac{2(\theta-1)}{1+n_{r}+N_{r}\mathcal{M}_{r}} \omega'(0) \;.
\end{align}

If $\theta=1$ is satisfied, then we recover the known isotropic case. Neglecting this case, we see that the condition is fulfilled when $\omega'(0)=0$.
%
%

%
\subsection{Case 2}
In this case, we make the following choices for the parameters: $\theta = 1$ and $N_{\perp} = N_{r}$. Thus, by simplifying the anisotropic term, we obtain
\begin{align}
    \Delta=\displaystyle P_{r 0}\left( \omega^{n_{r}+1} - \frac{1}{\omega^{N_{r}}}   \right)\left(\frac{\mathcal{M}_{\perp}-\mathcal{M}_{r}}{1-\mathcal{M}_{\perp}}\right).
\end{align}

Similar to the previous case, we perform the analysis of $\omega'$ at the origin. Thus, it can be proven that
\begin{align}
    \omega'(0)= \frac{2(1+n_{r}+N_{r})}{1+n_{r}+N_{r}\mathcal{M}_{r}} \left(\frac{\mathcal{M}_{\perp}-\mathcal{M}_{r}}{1-\mathcal{M}_{\perp}}\right) \omega'(0)\;,
\end{align}
which is also satisfied generically if $\omega'(0)=0$.
 %
\subsection{Case 3 }
Finally, by assuming $\mathcal{M}_{\perp} = \mathcal{M}_r$ and $\theta = 1$, we obtain a more complicated delta factor. To be concrete, $\Delta$ takes the form
\begin{align}
    \Delta = P_{r0} \, \mathcal{M}_{r}\left[ \frac{1}{\omega^{N_{r}}}-\frac{1}{\omega^{N_{\perp}}} \right].
\end{align}

In this final case, we observe that the condition $\omega'(0)=0$ is also maintained via the following equation:
\begin{align}
     \omega'(0)= \frac{2(N_{r}-N_{\perp})}{1+n_{r}+N_{r}\mathcal{M}_{r}} \omega'(0)\;.
\end{align}

It is important to note at this point that regardless of the case studied, taking the respective isotropic limit does not recover the known Lane--Emden equation because the equation of state for the radial pressures (Equation \eqref{pradial}) does not correspond to that of an ordinary polytrope. Thus, it is already possible to extend the modeled phenomenology by considering systems where $\Delta = 0$.

\begin{figure}[H]
\begin{adjustwidth}{-\extralength}{0cm}
\centering
\subfigure[]{\includegraphics[width=0.42\textwidth]{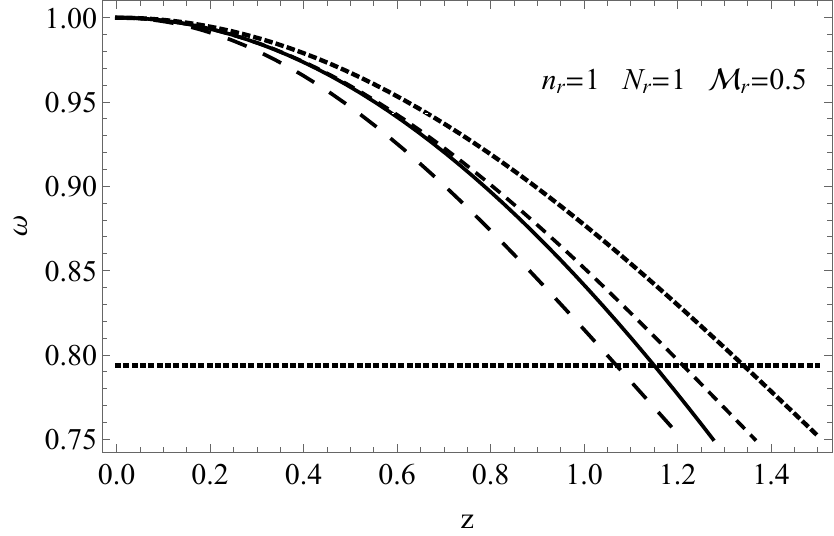}} \
\subfigure[]{\includegraphics[width=0.42\textwidth]{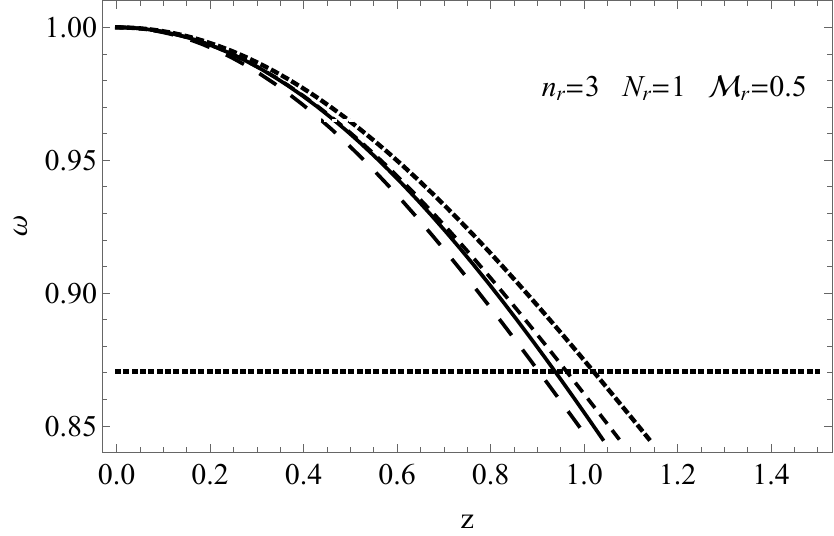}} \
\subfigure[]{\includegraphics[width=0.42\textwidth]{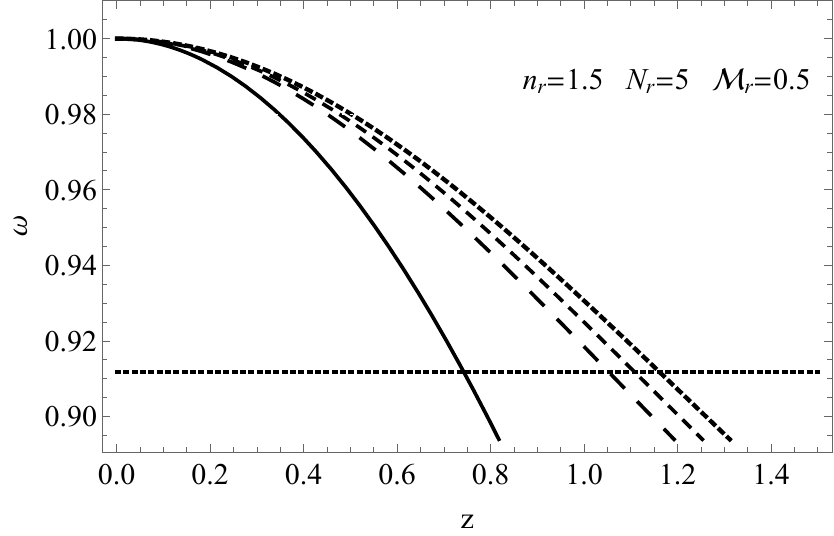}} 
\\
\subfigure[]{\includegraphics[width=0.42\textwidth]{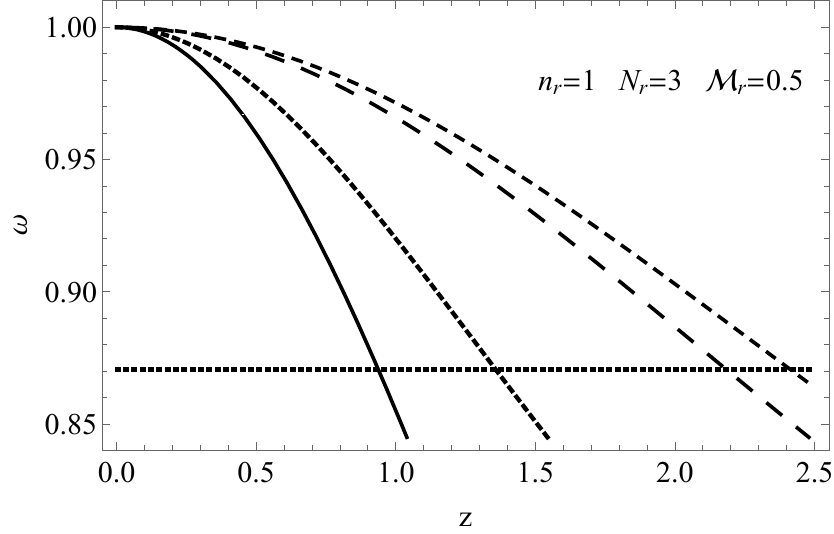}} \
\subfigure[]{\includegraphics[width=0.42\textwidth]{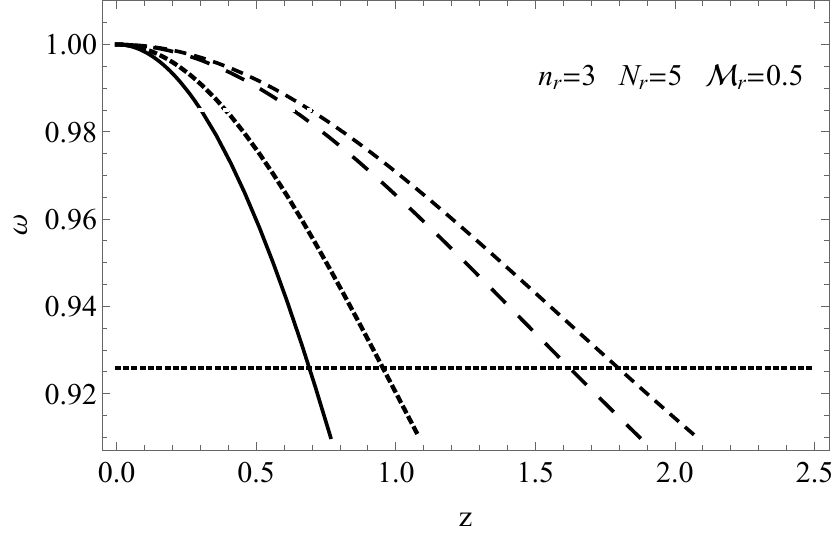}} \
\subfigure[]{\includegraphics[width=0.42\textwidth]{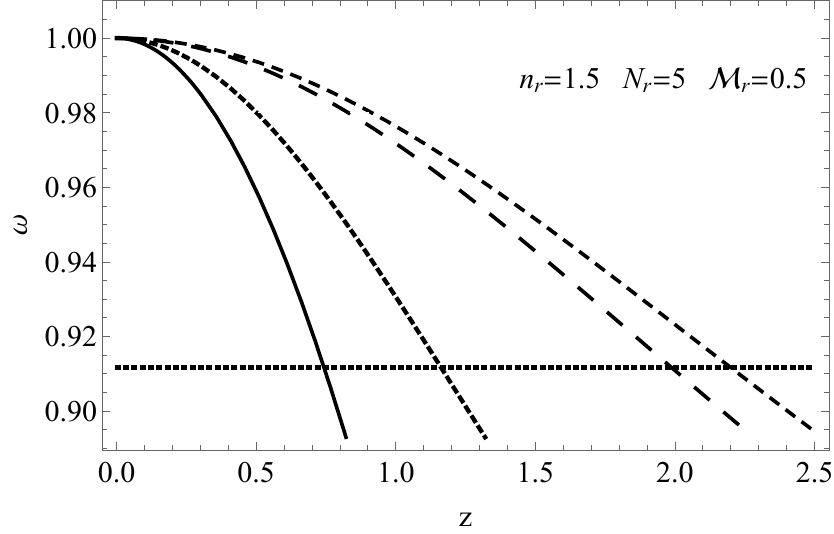}}
\\
\subfigure[]{\includegraphics[width=0.42\textwidth]{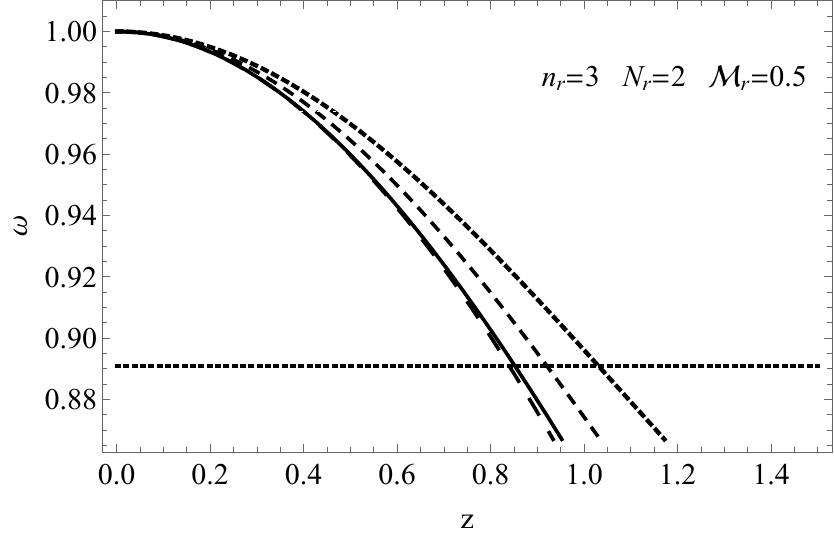}} \
\subfigure[]{\includegraphics[width=0.42\textwidth]{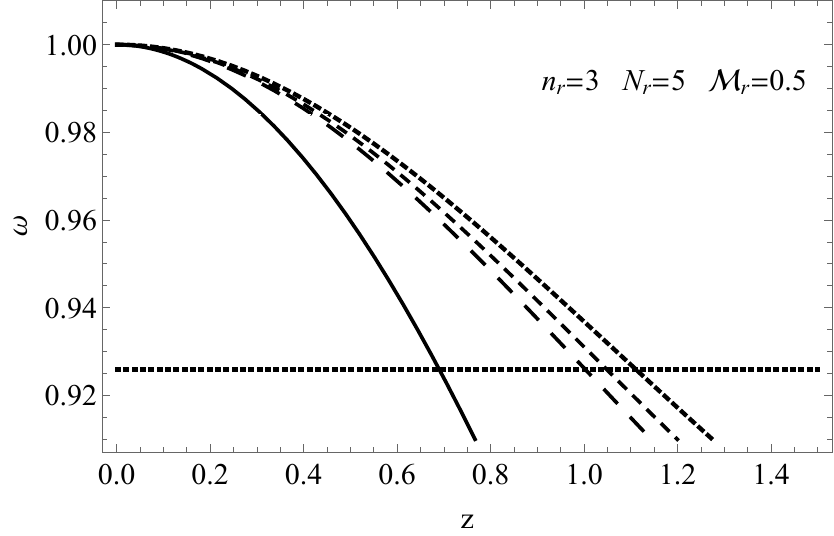}} \
\subfigure[]{\includegraphics[width=0.42\textwidth]{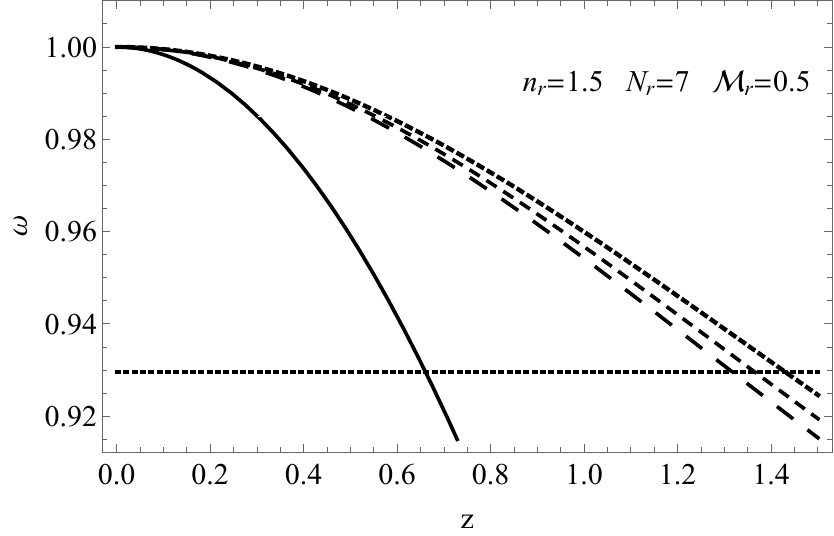}}
\end{adjustwidth}\vspace{-6pt}
\caption{Effective density versus radial coordinate. Each row corresponds to cases 1, 2 and 3. The isotropic Lane--Emden equation is shown with a solid line.
For case 1 (\textbf{a}--\textbf{c}),  
$\theta = 1$ (small dashed line), $\theta = 1.5$ (medium dashed line), and $\theta = 2$ (large dashed line).
For case 2 (\textbf{d}--\textbf{f}), $\mathcal{M}_{\perp} = 0.5$ (small dashed line), $\mathcal{M}_{\perp} = 2$ (medium dashed line), and $\mathcal{M}_{\perp} = 5$ (large dashed line).
For case 3 (\textbf{g}--\textbf{f}), 
$N_{\perp} = 1.5$ (small dashed line), $N_{\perp} = 3$ (medium dashed line), and $N_{\perp} = 4$ (large dashed~line).
}
\label{fig:1} 	
\end{figure}

{At this point, it is important to make some comments. In the subsequent discussion, we will examine one example for each case, amounting to a total of three examples. Our objective is to compare the anisotropic factor with other solutions in order to determine whether the solutions can accurately represent a well-defined anisotropic star. Consequently, we observe the following:
\begin{itemize}
\item Case 1: The anisotropic factor, denoted by $\Delta$, exhibited a negative value for the specific set of parameters utilized in the top-left panel of Figure \ref{fig:1}. This behavior aligns with other solutions found in the context of compact stars and is logical considering that the second term in the anisotropic formula becomes more significant than the first term. For further insights and related references, refer to \cite{Rincon:2023ens} and the cited sources therein.
\item Case 2: On this occasion, the anisotropic factor $\Delta$ showed a positive value for the set of parameters utilized in the middle-left panel of Figure \ref{fig:1}. Similar results were obtained in previous studies, which can be found in \cite{Bhar:2016lfu} for reference. In our current solution, the anisotropic factor increased for $\mathcal{M}_{\perp} = 5.0$ and $\mathcal{M}_{\perp} = 2.0$. However, when $\mathcal{M}_{\perp} = 0.5$, we observed that the isotropic case was recovered. This behavior is due to the fact that, according to Equation (38), when $\mathcal{M}_{\perp} = \mathcal{M}_r$, the anisotropic factor $\Delta$ becomes precisely zero.
\item Case 3: In this scenario, the anisotropic factor can either be positive or negative, depending on the specific numerical values considered. To be precise, when $N_{\perp} = 4.0$ and $N_{\perp} = 3.0$, the anisotropic factor decreases and becomes more negative. Conversely, when $N_{\perp} = 1.5$, the anisotropic factor takes on positive, well-defined values. This behavior can be attributed to the competition between the two terms involved in the definition of $\Delta$. Thus, when the first term is greater than the second term, $\Delta$ becomes greater than zero.
\end{itemize}

}
\vspace{-5pt}

\begin{figure}[H]
\begin{adjustwidth}{-\extralength}{0cm}
\centering
\includegraphics[width=0.42\textwidth]{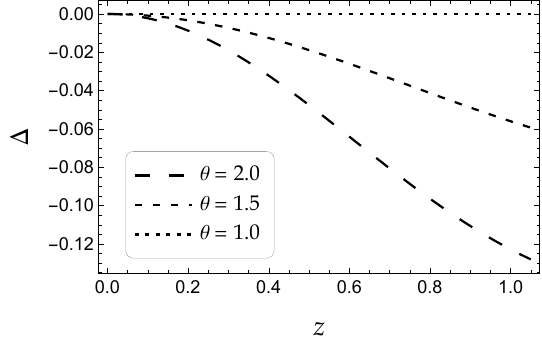} \
\includegraphics[width=0.42\textwidth]{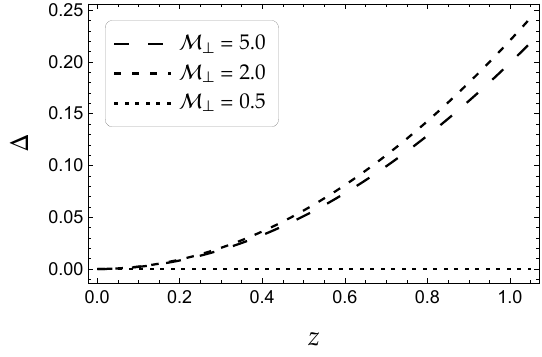} \
\includegraphics[width=0.42\textwidth]{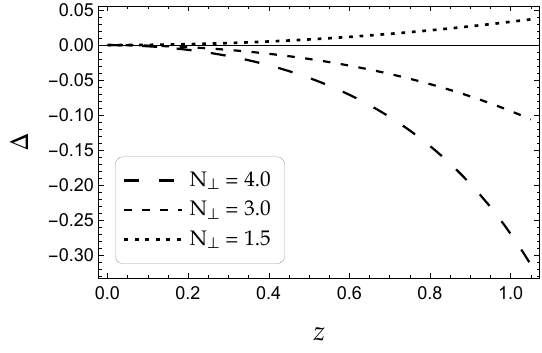} 
\end{adjustwidth}
\caption{Anisotropic factor 
 $\Delta$ against the coordinate $z$. The numerical values used to produce the figures are the following.
{(\bf{left panel})} We assumed the set $\{ \rho_c = 1, P_{r0} = 1, n_r = 1, N_r = 1,  \mathcal{M}_r = 0.5 \}$.
{(\bf{middle panel})}  We assumed the set $\{ \rho_c = 1, P_{r0} = 1, n_r = 1, N_r = 3, \mathcal{M}_r = 0.5, \theta = 1 \}$.
{(\bf{Right~panel})} We assumed the set $\{ \rho_c = 1, P_{r0} = 1, n_r = 3, N_r = 2, \mathcal{M}_r = 0.5 \}$.
}
\label{fig:3} 	
\end{figure}

In summary, by studying the anisotropy factor (see Figure \ref{fig:3}) and comparing our results with other work, we can confirm that our model could be used to describe anisotropic stars, given the similarity of our results. A more in-depth and robust analysis will be carried out when the relativistic solution is studied.

\section{Chandrasekhar Mass}
As a consequence of what has been developed so far, it is interesting to evaluate how the generalized Chaplygin equation of state and the proposed anisotropy modify the Chandrasekhar mass. The total mass of the corresponding distribution is given by the well-known expression 
\begin{equation}\label{mass}
    M \;=\; 4\pi \int_0^R \! r^2\rho \,dr \;=\; 
    4\pi \rho_c \frac{R^3}{(z_0)^3} \int_0^{z_0} \! z^2\omega^{n_r} \,dz   \;,
\end{equation}
where the value $z_0$ satisfies the boundary pressure condition in Equation \eqref{p-cond}.
\noindent
Using the expression in Equation \eqref{lanenew}, and after a lengthy but straightforward calculation, we arrive at the following simplified form:
\begin{eqnarray}\label{mass-pre}
    z^2 \omega^{n_r} =
    \frac{d}{dz}\Bigg[ 
    -(z^2 \alpha \omega') +
    \frac{2}{P_{r0}(1+n_r)}\frac{z}{\omega^{n_r}}\Delta
    \Bigg], \;\;\;
\end{eqnarray}
with $\alpha$ given by Equation \eqref{alpha}. Note that for $\Delta=0$, an expression formally equal to the usual case is obtained, but this time, $\alpha\neq 1$ must be considered, which makes the mass expression receive contributions due to the generalized equation of state.

\noindent
By substituting the expression in Equation (\ref{mass-pre}) into Equation (\ref{mass}), we can find an equation for the total mass relative to the Chandrasekhar mass. This is
\begin{eqnarray}\label{mass-ratio}
    \frac{M}{M_{Ch}} = \Big(\frac{z_{ch}}{z_0} \Big)^3
    \Bigg[
    \frac{-(z^2_0 \alpha_0 \omega'_0) + \displaystyle
    \frac{2}{P_{r0}(1+n_r)}\frac{z_0}{\omega^{n_r}_0}\Delta_0}{-z_{ch}^2 \omega'_{ch}}
    \Bigg] , \nonumber \\
\end{eqnarray}
where all quantities $f_i$ are evaluated at the corresponding outer radius $z_i$ where pressure vanishes (i.e., $f_i=f(z_i)$). It is remarkable that the expression in Equation \eqref{mass} has an exact analytical expression that can be evaluated and contrasted with the well-known expression for the Chandrasekhar mass.

\section{Concluding Remarks}\label{final}

In this work, we described a complete family of Chaplygin-type equations of state for anisotropic matter. We assumed that both the radial and tangential pressures had an equation of state as described previously. This has the advantage that, at least for small deviations from the isotropic case, the model is well justified. This differs from the type of anisotropy model discussed in \cite{Herrera:2013dfa}.

We derived the Lane--Emden equations that defined the model. The different parameters allowed us to particularize some cases that we found to be relevant. We then applied a numerical treatment to solve the equations arising in each case, calculated the total mass and compared it with the Chandrasekhar mass. It is clearly shown that the Chandrasekhar mass limit changed with the introduced anisotropy. These models can be further developed and used to study the influence of local anisotropy on such important problems as the Chandrasekhar mass limit, particularly in relation to the possible origin of super-Chandrasekhar white dwarfs. We find it relevant to further explore these models and to try to determine whether the super-Chandrasekhar white dwarfs inferred from the data gathered \cite{howell2006,scalzo2010,scalzo2012,hachisu2012,das2013} are the result of models, as considered in this work. This interesting question deserves more attention in future work.

\begin{figure}[H]
\begin{adjustwidth}{-\extralength}{0cm}
\includegraphics[width=0.65\textwidth]{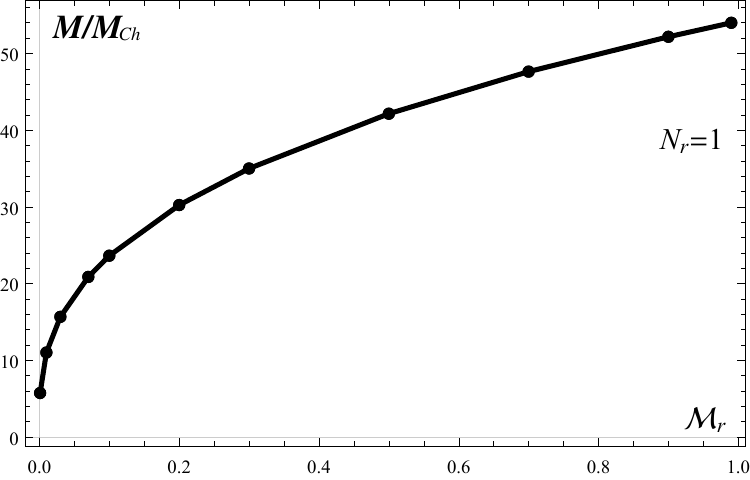} \
\includegraphics[width=0.65\textwidth]{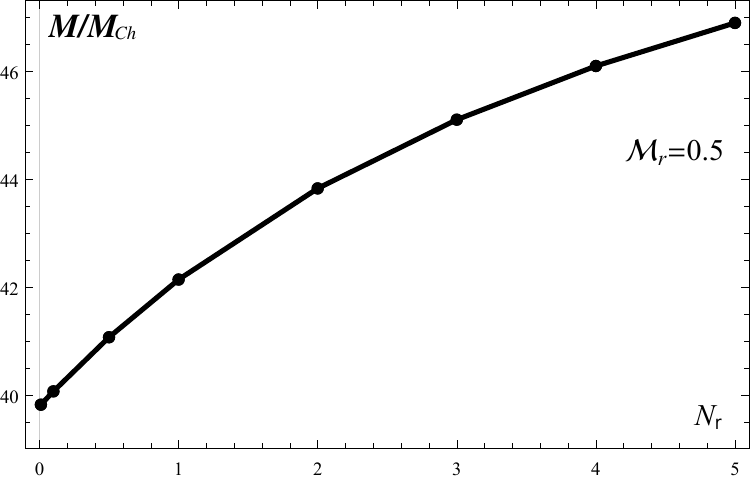} 
\\
\includegraphics[width=0.65\textwidth]{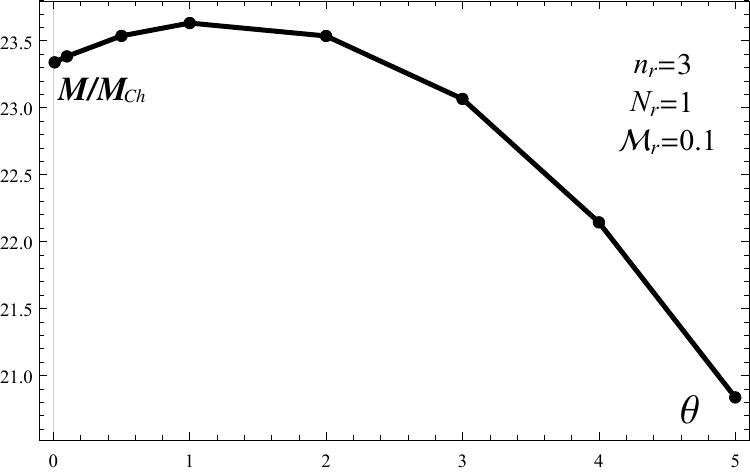} \
\includegraphics[width=0.65\textwidth]{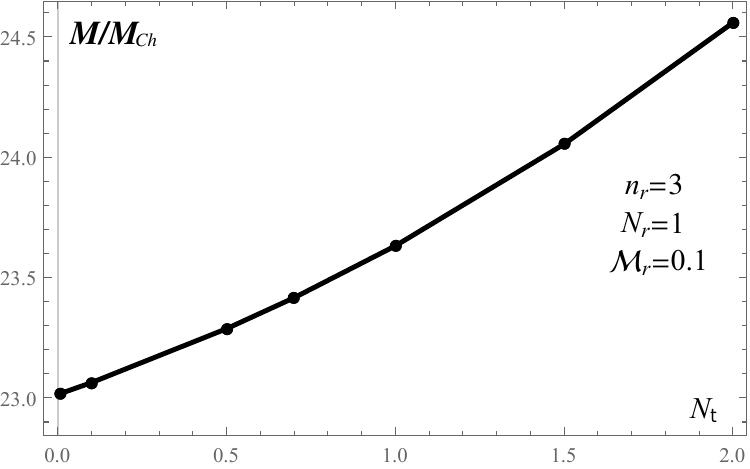} 
\end{adjustwidth}
\caption{Total mass relative to Chandrasekhar mass. The top row corresponds to two isotropic cases: (\textbf{left panel}) when $\mathcal{M}_r$ varies and  (\textbf{right panel}) when $N_r$ varies. The bottom row corresponds to two of the anisotropic cases. The left panel is for anisotropic case 1, and the right panel is for anisotropic case 3.
}
\label{fig:2} 	
\end{figure}

Looking at the plots in Figure \ref{fig:1}, we can appreciate how different sets of parameters affected the solution. In particular, we see that it is possible to have both expansion and contraction with respect to the isotropic solution, depending on the case considered. Cases 2 and 3 seemed to cause stronger deviations even with small changes in the parameters. When comparing how the mass is modified by including the generalized equation of state with respect to the Chandrasekhar mass, it can be seen that, in general, more mass can be captured. This effect was increased by including the effects of anisotropy, as can be seen from the graphs in Figure \ref{fig:2}. These effects can be used to model some phenomenological aspects of compact objects and even galactic scales.

\vspace{3pt}

At this point, it is necessary to mention the following. While it is usual to compare the numerical solution with observational data, such a comparison is not convenient in this case. There are observational constraints in the context of stars, mainly in the case of relativistic compact stars. Such restrictions (mass and radius) are commonly compared with the mass-radius profile of a concrete problem, imposing limitations on the mass and radius of a star. Given that these constraints are well established in the full relativistic case, and as we were only considering the Newtonian case here, we decided to conduct a complete analysis, including the relativistic case, non-relativistic case and a comparison with another compact star in future work. Having said that, there are studies that compare both relativistic and non-relativistic solutions \cite{Oliveira:2014mka,Olmo:2019flu}. Some of these studies have shown that the mass-radius profile in the Newtonian case predicts a more massive and larger star. This is consistent with the conventional approach of ignoring pressure in the Newtonian case. 
Lastly, it is crucial to highlight, in accordance with a recent paper \cite{Pretel:2023nhf}, that although we did not place constraints on a Newtonian dark star, the relativistic scenario may offer some insights into how anisotropy influences the evolution of such stars. As emphasized in \cite{Pretel:2023nhf}, if a star becomes increasingly anisotropic, then it permits a substantial augmentation in the maximum mass, thereby providing a more favorable explanation for the observed compact objects in nature.

Finally, it should be emphasized that this work was developed in the context of Newtonian gravity using spherical symmetry. It is possible that this symmetry can be broken by the same kind of physical factors that create the anisotropy in the system. In this case, the method described here should be applied with some caution and only as an approximation.



\vspace{6pt}

\authorcontributions{Conceptualization, G.A., \'A.R. and E.S.; 
methodology, G.A., \'A.R. and E.S.; 
formal analysis, G.A., \'A.R. and  E.S.; 
investigation, G.A., \'A.R. and E.S.; 
writing---original draft preparation, G.A., \'A.R. and E.S.; 
visualization, G.A., \'A.R. and  E.S.  
All authors have read and agreed to the published version of the manuscript.
}

\funding{\hl{ }}

%

\dataavailability{\hl{ }} 

\acknowledgments{We thank the anonymous reviewer for constructive criticism and useful comments that improved the quality of the manuscript.
A.R. is funded by the Mar{\'i}a Zambrano contract ZAMBRANO 21-25 (Spain).
A.R. acknowledge financial support from the Generalitat Valenciana through PROMETEO PROJECT CIPROM/2022/13.
}

\conflictsofinterest{The authors declare no conflict of interest.}

\begin{adjustwidth}{-\extralength}{0cm}


\reftitle{References}



\PublishersNote{}
\end{adjustwidth}

\end{document}